# Normal State magneto transport properties of $FeSe_{0.5}Te_{0.5}$ superconductor: The role of topological surface state

N. K. Karn[1,2], M. M. Sharma[1,2], & V.P.S. Awana[1,2,*]

[1]*Academy of Scientific & Innovative Research (AcSIR), Ghaziabad, 201002, India.*

[2]*CSIR- National Physical Laboratory, New Delhi, 110012, India.*

**Abstract:**

Doped Iron Chalcogenide (FeCh) superconductors are extensively studied in the context of topological superconductivity. However, the evidence of topological surface states in electrical transport measurements of the doped FeCh system is yet warranted. In the present letter, we report the successful growth of the single crystal of $FeSe_{0.5}Te_{0.5}$, with the growth direction along the c-axis confirmed by single-crystal X-ray diffraction and High-resolution Transmission electron microscopy (HRTEM). The angle-dependent magneto transport measurements have been performed on a single crystal of doped FeCh system, i.e., $FeSe_{0.5}Te_{0.5}$. A non-saturating linear magnetoresistance (MR) has been observed under the magnetic field up to ±14 T (Tesla) in the normal state of $FeSe_{0.5}Te_{0.5}$. The MR is shown to possess anisotropy, which indicates the presence of topological surface states in $FeSe_{0.5}Te_{0.5}$. Angle- dependent Magneto-conductivity (MC) at low magnetic fields has been modelled by Hikami Larkin Nagaoka (HLN) formalism, which shows the presence of weak antilocalization (WAL) effect in $FeSe_{0.5}Te_{0.5}$. The observed WAL effect is found to be 2D in nature through angle-dependent magneto transport measurements. Theoretical calculations based on Density Functional Theory (DFT) are also performed to get more confidence on the presence of topological surface states in $FeSe_{0.5}Te_{0.5}$.

*Corresponding Author
Dr. V. P. S. Awana
Chief Scientist,
CSIR-NPL, New Delhi, India, 110012
E-mail: awana@nplindia.orgPh. +91-11-45609357,
Fax-+91-11-45609310
Homepage: www.nplindia.org/wp-content/uploads/common/awana@nplindia.pdf

## Introduction

Superconducting materials with topological surface states are supposed to be crucial materials to realize an ideal candidate for topological superconductivity [1-3]. An ideal topological superconductor can provide suitable conditions for the emergence of Majorana-bound states [3,4]. Doped FeCh superconductors are extensively studied for their remarkable superconducting properties. Monolayer of FeSe deposited on STO substrate is reported to show a high $T_c$ as 100K [5]. Also, the scanning tunneling spectroscopy (STS) and angle-resolved photoelectron spectroscopy (ARPES) measurements performed on the FeSe monolayer show the existence of 1-D topological edge states [6,7]. Apart from monolayer FeSe, topological states are also observed in Te doped FeSe system, as doping of Te atoms in the FeSe unit cell enhances the system's effective spin-orbit coupling (SOC). The doped FeCh system in which Te atom doping is near 50% has emerged as a potential candidate to observe the simultaneous existence of non-trivial topological properties and superconductivity [8-10]. There has been significant evidence of the topological nature of $FeSe_{1-x}Te_x$ on theoretical fronts [8-12]. Experimentally, Majorana bound states are observed in STS measurements performed inside the vortex core of the doped FeCh system, i.e., $FeSe_{1-x}Te_x$ [13,14]. ARPES measurements performed on $FeSe_{1-x}Te_x$ have confirmed the presence of the Dirac cone [15,16]. In a particular doping range, i.e., $0.55<x<0.7$, the observed Dirac cone becomes gapped in superconducting state at the fermi level and at the Dirac point, showing that time reversal symmetry is broken in the superconducting state due to surface ferromagnetism [16].

Though the system $FeSe_{1-x}Te_x$ is extensively studied for its superconducting and exotic topological properties [8-18], a detailed report on normal state transport properties is still missing particularly for x=0.5, as per our knowledge. There have been reports on transition metal doping in FeCh system leading to different phases [19-21]. Normal state transport properties of topological materials are greatly influenced by topological surface states and cause the emergence of weak antilocalization (WAL) effect [22,23], non-saturating linear magnetoresistance (MR) [24,25], quantum oscillations [26-29], etc. Different mechanisms have been proposed to explain the linear component of the magnetoresistance in FeCh system, Licciardello et al. [30] show coexistence of orbital and quantum critical magnetoresistance in $FeSe_{1-x}S_x$, Wang and coworkers suggest that spin fluctuations are the origin of the linear magnetoresistance in FeSe films [31], while Amige et al. invoke the multiband character of the transport for the observed behavior in MR of FeSe [32].

In this letter, we report the growth of single crystalline $FeSe_{0.5}Te_{0.5}$, which is well characterized by X-ray diffractometer (XRD) and Selected Area Electron Diffraction (SAED) patterns for crystallinity and the phase purity, the microstructural and morphological character is elucidated through scanning electron microscopy (SEM) and High-resolution Transmission Electron Microscopy (HRTEM). Further, the normal state transport properties of $FeSe_{0.5}Te_{0.5}$ through angle-dependent magneto-transport measurements. A non-saturating linear MR has been observed in $FeSe_{0.5}Te_{0.5}$. Low magnetic field transport properties hint towards the presence of the WAL effect, which is confirmed through HLN-fitted magnetoconductivity. The origin of the observed WAL effect is probed through angle-dependent magneto-transport measurements, which clarifies that topological surface states mainly dominate the normal state transport properties. Also, the presence of topological surface states is confirmed through the surface state spectrum of $FeSe_{0.5}Te_{0.5}$ using the first principle methods. This letter will prove to be a good addition to the literature demonstrating the existence of topological surface states in $FeSe_{0.5}Te_{0.5}$.

## Experimental

Density functional theory (DFT) has been used to simulate the electronic band structure and find the ground state electronic density. The calculation is implemented in Quantum Espresso [33,34], where PBE-type functionals with generalized gradient approximation GGA are used to inculcate the corrections due to electronic exchange and correlations [35]. For SOC based calculations, PBE functionals with full-relativistic corrections are used from the PSEUDODOJO library of functionals. The k-space is sampled on a mesh of 12×12×7 Monkrost-Pack [36] with charge cutoff 1080 Ry, wfc cutoff is set to 90 Ry, degauss is 0.015, total energy convergence threshold is $6\times10^{-5}$ Ry and SCF electronic convergence cutoff is set to $1.2\times10^{-9}$ Ry. Further, the 26 calculated wavefunctions are wannierised in Wannier90 [37] with convergence tolerance $10^{-7}$ a.u. and implemented in WannierTools [38]. The tight binding model obtained from Wannier90 is solved iteratively using Green's function method [39, 40] to get the surface state spectrum.

The single crystal of $FeSe_{0.5}Te_{0.5}$ is synthesized using a solid-state reaction route based on well-optimized heating protocols as described in ref. 18. The constituent elements Fe, Te and Se are mixed in Ar gas-filled glove box in the stoichiometric ratio 1:0.5:0.5. The homogenized mixture is palletized and sealed in vacuum-encapsulated quartz tube at $10^{-5}$ torr pressure. The encapsulated pallet is subjected to heat treatment in an automated furnace. The sealed quartz tube is heated to $450^0$ C in 4 h and kept for 4 h and then the temperature is increased to $1000^0$ C in 6 h and dwelled for 24 hours. Finally, the furnace is slowly cooled ($4^0$ C/hour) down to room temperature. Rigaku mini flex tabletop XRD equipped with Cu-K$_\alpha$ radiation of wavelength 1.54 Å is used to record the XRD pattern of the synthesized FeSe$_{0.5}$Te$_{0.5}$. The HRTEM and SAED pattern is recorded on JEOL 2100F at 1 nm and 2 nm$^{-1}$ resolution, respectively. Image-J software is used to analyze the observed data. The SEM and energy dispersive X-ray analysis (EDAX) measurements were performed on Zeiss EVO-50 to examine the surface morphology and elemental composition of as synthesized crystal. Further, vibrational modes were studied by recording Raman spectroscopy on Renishaw in Via Reflex Raman Microscope equipped with 532 nm laser. Electrical transport measurements have been carried out using a Quantum Design (QD) made Physical Property Measurement System (PPMS) equipped with a horizontal sample rotator. A standard four-probe method has been used for the magneto transport measurements, in which four linear contacts with silver epoxy have been made on the surface of the sample. The sample is mounted on the resistivity puck keeping the growth axis parallel to the magnetic field and the current is perpendicular to the applied field. As the sample is rotated, the applied magnetic field direction not only changes with the c-axis (growth axis) but also changes with the current direction as well.

## Results and Discussion

In order to understand the semi-metallic behavior of FeSe$_{0.5}$Te$_{0.5}$, the DFT calculations are performed to calculate the electronic band structure and orbital projected density of states. Fig. 1(a) shows the high symmetric path Γ-X-M-Γ-Z-R-A-Z chosen for the calculation of bulk band structure. Fig. 1(b) shows the calculated iso-energy Fermi surface corresponding to those five bands nearby and crossing the Fermi level. Out of these five, three surfaces cross the first Brillouin Zone (FBZ) boundary largely contributing charge carriers. The rest two surfaces cross the FBZ in the k$_z$ direction only. The calculated band structure is shown in Fig. 1(c). It shows the bands calculated using SOC and without SOC effects. We find that there are five bands crossing the Fermi level, indicating ample charge carrier for conduction. In calculations, DFT-D3 [41] corrections are also included to incorporate the effects of Vander Waals gap present in FeSe$_{0.5}$Te$_{0.5}$. The VdW correction energy is found to be -0.06 Ry. From, the calculation we find that the effect of SOC is moderate as one can see band splitting due to SOC especially along the path Γ-Z. The inset of Fig. 1(c) shows the enlarged band structure along Γ-Z path. Just behind Γ point there are three without SOC bands, which split into six non-degenerate bands when SOC is included. Such non-degenerate band splitting confirms the absence of inversion symmetry. A similar non-degenerate band splitting due SOC has been found in AuSn$_4$ for its non-Centro-symmetric phase [42]. The unit cell shown in the inset Fig 3(a) also shows FeSe$_{0.5}$Te$_{0.5}$ lack inversion symmetry as Se and Te sites are inversion points occupied with different atoms. The presence of SOC can be associated with the heavier atoms i.e. the orbital angular momentum of these atoms is moderate. The Te atom is the heaviest one, which might induce SOC but, later we find that near the Fermi level bands, the contribution from the p-orbital of the Te atom is minimal. Due to this, the observed effects of SOC i.e. band splitting are significantly observed slightly away from the Fermi level.

Fig. 1(d) shows the calculated projected density of states (PDOS) without SOC. It shows that the bands originating near the Fermi level are mostly from the d-orbital of Fe and p orbitals of Te and Se atoms. The total DOS plot shows a non-zero value (sufficiently large) of DOS at the Fermi energy indicating the expected metallic behavior of FeSe$_{0.5}$Te$_{0.5}$. We also observe that in total DOS, the d-orbital contribution of Fe is quite large whereas the p-orbitals of the rest atoms are minimal.

From the band structure calculation, we got an indication of the presence of a type-II Dirac cone in FeSe$_{0.5}$Te$_{0.5}$. Therefore, we calculated the surface state spectrum of the material by wannierising the Bloch wavefunction obtained from DFT. From there we constructed a tight-binding model and solved it iteratively using Green's function method to get to the surface state spectrum. The surface card is set into the (001) plane i.e. the chosen path for the surface state spectrum is X-Γ-M in $k_z = 0$ plane. The calculation is done on a much-dense grid, 81 slices of one reciprocal vector.

Fig. 2 shows the surface state spectrum of $FeSe_{0.5}Te_{0.5}$. At Γ point, 0.25 eV above the Fermi level, a Dirac cone type-I shape is seen but the inverted cusp is not sharp. However, near this point, close to 0.1 eV along both sides i.e. X-Γ and Γ-M, type-II Dirac cones are observable as marked in Fig. 2. Here, the observed Dirac point at Γ point is above the Fermi Level, which is not in agreement with the experimental results obtained from ARPES measurement on $FeSe_{0.5}Te_{0.5}$ [15, 16]. ARPES results show the Dirac cones are found below the Fermi level. It is a known mismatch of experimental and computational results [15, 16]. The studied $FeSe_{0.5}Te_{0.5}$ system is reported to have multiple topological states [15]. For topological insulating materials, a Dirac cone is expected to be observed below or at the Fermi Level, while for some of the topological semimetal, an extra Dirac cone is seen above the Fermi Level which is termed as topological Dirac semimetal (TDS) cone [15, 43]. Here, in the studied $FeSe_{0.5}Te_{0.5}$ system, the observed Dirac cone is above the fermi level and can be termed as TDS cone as observed for topological semimetals [15, 43].

Now coming to experimental results, after heat treatment, the obtained crystals are silvery shiny and around 1cm long as shown in the left-hand side inset of Fig. 3(a). Fig. 3(a) exhibits the Rietveld refined powder XRD (PXRD) pattern of synthesized $FeSe_{0.5}Te_{0.5}$ crystals. $FeSe_{0.5}Te_{0.5}$ is crystallized in a tetragonal phase with P4/nmm space group. The goodness of fit parameter, i.e., $\chi^2$ is found to be 1.56, and the obtained lattice parameters are $a=b=3.789(8)$ Å and $c=5.930(7)$ Å, and $\alpha=\beta=\gamma=90°$. The results of PXRD refinement are summarized in Table 1. All observed peaks in the PXRD pattern are indexed with their respective plane. Here, for the synthesized $FeSe_{0.5}Te_{0.5}$, no sign of the presence of the impurity phase of -FeSe ($Fe_7Se_8$) has been observed as reported for pure FeSe crystals [44, 45]. The obtained lattice parameters are in accordance with the previously reported values [18] and are in accordance with those obtained for pure FeSe crystal [44, 45]. The unit cell is expanded along the c-axis due to comparatively larger Te atoms doping on the Se site in the FeSe unit cell [44, 45]. The inset in the middle of Fig. 3(a) shows the VESTA-drawn unit cell of synthesized $FeSe_{0.5}Te_{0.5}$ crystals, which indicates that both Te and Se atoms occupy different atomic sites in unit cells. To confirm the unidirectional growth of synthesized $FeSe_{0.5}Te_{0.5}$, the XRD pattern is recorded for the mechanically cleaved crystal flake, and the same is shown as the right-hand side inset of Fig. 3(a). The plot depicts the presence of sharp and intense XRD peaks only for specific Bragg's positions, which corresponds to (0 0 1), (0 0 2), (0 0 3) and (0 0 4) planes. These show that the crystal exhibits unidirectional growth along the c-axis, thus confirming the single crystalline nature of the synthesized $FeSe_{0.5}Te_{0.5}$ sample.

Further, the SAED pattern of the synthesized $FeSe_{0.5}Te_{0.5}$ is recorded and shown in the left image in Fig. 3(b). This image reveals symmetrically positioned bright spots, providing evidence of the crystalline structure in the synthesized crystal. Notably, the spots corresponding to (0 0 2), (0 0 3), and (0 0 4) are visibly marked in the image. The right-hand side of Fig. 3(b) depicts the HRTEM image recorded at 1 nm resolution which illustrates the staking of (0 0 2) planes. The calculated interplanar spacing for the (0 0 2) plane is determined to be 2.96 Å, which closely aligns with the value obtained from the XRD pattern.

Fig. 3(c) shows the EDAX spectra with the inset table showing the elemental atomic and weight percentage of the constituent elements. The observed peaks confirm the absence of foreign elements. From atomic weight percentage, the calculated stoichiometry of the sample is $FeSe_{0.44}Te_{0.5}$ which is within the uncertainties close to $FeSe_{0.5}Te_{0.5}$. The SEM image as shown in Fig. 3(d), depicts the surface morphology of the as-grown sample. The single crystalline nature with layered growth along the direction c-axis is endorsed by the SEM image visualized at 5 μm resolution. The single-phase growth of the crystal is evidenced by the absence of other contrasting colors in the SEM image.

Fig. 3(e) shows the recorded Raman spectroscopy of as grown sample $FeSe_{0.5}Te_{0.5}$ to trace out the vibrational modes present at room temperature. In the recorded data, peaks have been identified, fitted and analyzed using the Lorentz deconvolution method as shown in Fig. 3(e). The observed peaks in the Raman shift are in accordance with the previous reports [46, 47]. The discrepancy in assigning vibrational modes with corresponding peaks in previous works is well highlighted by Lopes et al [46]. For the tetragonal systems FeSe/Te, theoretically, four Raman active modes are predicted by Xia et. Al. [48], where they show the Raman peak position is well dependent on the lattice parameters. Here, for FeSe$_{0.5}$Te$_{0.5}$, five identified Raman peaks are observed at 81.4 ± 0.53, 129.3 ± 0.1, 145.9 ± 0.14, 179.4 ± 2.03 and 269.6 ± 5.03 cm$^{-1}$. The first three peaks are attributed to low-frequency modes E$_g$, A$_{1g}$ and B$_{1g}$ following [46, 47]. The deviation in peak positions from ref. 46 can be attributed to the difference in the exact lattice parameters of the as-grown sample.

The inset of Fig. 4(a) shows the $\rho$-$T$ plot of FeSe$_{0.5}$Te$_{0.5}$, confirming the presence of superconducting transition at around 13.5 K. The obtained resistivity plot is similar to as reported in ref. 37. Superconducting properties of FeSe$_{0.5}$Te$_{0.5}$ are well elucidated in the literature [18, 49]. Here, normal state transport properties are studied to get magneto transport evidence of the non-trivial topological character of FeSe$_{0.5}$Te$_{0.5}$. Fig. 4(a) shows the MR% vs. the applied magnetic field ($H$) plot of FeSe$_{0.5}$Te$_{0.5}$ at different temperatures viz. 15, 20, 30, and 40 K. The magnetic field is varied in a range of ±14 T. MR data has been taken in for both positive and negative magnetic fields, and the mean $(MR_{(+H)} + MR_{(-H)})/2$ is taken to uphold the symmetry of the plot, here $MR_{(+H)}$ showing the MR values in a positive magnetic field and $MR_{(-H)}$ is showing the MR values in a negative magnetic field. This mean value of MR is plotted against the applied magnetic field and the same is shown in Fig. 4(a). FeSe$_{0.5}$Te$_{0.5}$ exhibits a linear non-saturating MR of around 0.8% at 15 K. Though the value of MR is very low, it is comparable to other topological materials [50-53]. The value of MR% is lowered as the temperature is raised, and the same is almost suppressed at 40K. MR% curve sharply increases at lower magnetic fields, which indicates that the WAL effect may be present in the system. In topological materials, the WAL effect results in a sharp increment in MR% at low magnetic fields in the V-shape [52]. This V-type MR% at low magnetic fields broadens with increments in temperature as the contribution due to surface states weakens. WAL effect can emerge due to the intrinsic high SOC of the material or due to the topological surface states. To determine the origin of the observed WAL effect, angle-dependent magneto transport measurements have been carried out in the normal state of the synthesized FeSe$_{0.5}$Te$_{0.5}$ single crystal.

Fig. 4(b) shows the resistivity with respect to the tilt angle in the presence of the magnetic field of 14 T. Here the tilt angle is measured from the z-axis, and the geometry used for the angle-dependent magneto transport measurements has been shown in the inset of Fig. 4(b). Here resistivity shows strong anisotropy when the sample is tilted with respect to the applied magnetic field. The topological surface states dominated transport properties are strongly dependent on the transverse component of the magnetic field and it is suppressed as the orientation deviates from the transverse one [53-55]. Here, the resistivity is found to be maximum when the magnetic field is perpendicular to the applied current and the same is minimum when both are parallel. The resistivity plot is fitted with the equation $\rho(\theta) = |A + B\cos(\theta)|$, here A and B are fitting constants. The resistivity is found to satisfy the equation. It shows that topological surface states dominate the transport properties at 15 K. Further, the MR% has been measured at 15 K for various tilt angles viz. 0º, 15º, 30º, 45º, 60º, 75º and 90º and the same is shown in Fig. 4(c). MR% also shows an apparent anisotropy with the tilt angle by showing the highest MR at 0º and the lowest at 90º, indicating the possibility of topological surface states dominating the transport mechanism. It is well known that the topological surface states dominated transport is very susceptible to magnetic field direction, though the contribution of bulk conducting states is considered to be isotropic [55]. For parallel magnetic fields, the effect of topological surface states on transport properties is considered to be suppressed entirely. Here, to determine the contribution of topological surface states in the observed WAL effect the bulk state contribution has been subtracted from using the following relation

$$\Delta\sigma_{\theta,B} = \frac{l}{w}\left(\frac{1}{R_{\theta,B}} - \frac{1}{R_{90°,B}}\right) \tag{1}$$

Here, $l$ is the length between the voltage contacts and $w$ is the width of the sample. $\Delta\sigma_{\theta,B}$ is longitudinal sheet magnetoconductance. $\Delta\sigma_{\theta,B}$ at 15K for all measured angles has been plotted against $H\cos\theta$ is Fig. 4(d) in low magnetic field range. All curves are found to coincide with each other suggesting that the observed WAL effect is 2D in nature and is governed by topological surface states. The coincided curves are also fitted with HLN equation [55-57], which is as follows:

$$\Delta\sigma(H) = \frac{-\alpha e^2}{\pi h}\left(ln\left(\frac{B_\varphi}{|H\cos\theta|}\right) - \Psi\left(\frac{1}{2} + \frac{B_\varphi}{|H\cos\theta|}\right)\right) \quad (2)$$

Here, $\Delta\sigma$ represents the difference between field-dependent MC and zero-field MC. $B_\varphi$ denotes the characteristic magnetic field and is related to phase coherence length $l_\varphi$ with the relation $B_\varphi = h/(8e\pi l_\varphi^2)$. The obtained fitting parameters are found to be $\alpha = -0.44$, and $l_\varphi = 59.87$ nm. The negative value of suggests that the WAL effect is present in the system. Also, the value of signifies the contribution of conducting channels in the conduction mechanism. The standard values of for topological states governed conduction phenomenon are -0.5 (single topological state) and -1.0 (two distinct topological states). Any deviation from these standard values signifies that the conduction mechanism is contributed by both topological surface and bulk conducting states [51-56]. These values are very close to the previous fitting results. This value of $\alpha$ is very close of standard value of -0.5, and suggest that the observed 2D WAL effect is mainly contributed by single surface state. The SOC effect plays key role in the phenomena like WL/WAL or even in the presence of topological surface states [58]. In FeSe$_{0.5}$Te$_{0.5}$ also, the HLN analysis of magneto-transport suggest that the system under investigation has SOC interaction and the same is supported by DFT band structure calculations.

## Conclusion

Summarily, in this letter, a systemic study has been performed on normal state transport properties of a superconducting single crystal of FeSe$_{0.5}$Te$_{0.5}$. The synthesized single crystal of FeSe0.5Te0.5 is well characterized by XRD and HRTEM/SAED showing the single crystal growth along the c-axis. The SEM/EDAX measurement confirms layered growth in with all elements in stoimetry. The normal state of FeSe$_{0.5}$Te$_{0.5}$ exhibits a non-saturating linear MR, which is accompanied by WAL effect. The observed WAL effect is examined through anisotropic magneto transport properties. $\rho$ vs $\theta$ curve shows a clear anisotropy, which hints towards topological states dominated WAL effect. This is further clarified by measuring longitudinal sheet magneto conductance for different tilt angles, which are plotted against $H\cos\theta$. All curves merged into each other, and are found to be well fitted with HLN equation, which pointed out that the studied FeSe$_{0.5}$Te$_{0.5}$ single crystal exhibit 2-D WAL effect, which is mainly governed by a single topological surface state. Also, some type-II Dirac cone-like features are observed in DFT calculated bulk electronic band structure, which also evidenced the presence of topological surface states in FeSe$_{0.5}$Te$_{0.5}$.

## Acknowledgment:

Authors would like to thank Director of National Physical Laboratory, New Delhi for his keen interest. Authors acknowledge Dr. J. Tawale and Ms. Sweta for SEM/EDAX and Raman spectroscopy measurements, respectively. Authors also acknowledge Prof. S. Patnaik for the HRTEM/SAED measurements. N. K. Karn and M.M. Sharma would like to thank CSIR, India for research fellowship and AcSIR, Ghaziabad for Ph.D. registration. This work is supported by in house project number OLP 240832 and OLP 240232.

## Data Availability Statement:

All the data associated to the MS will be made available on reasonable request.

## Declaration:

The authors declare that they have no known competing financial interests or personal relationships that could have appeared to influence the work reported in this paper.

## Author Contribution
N. K. Karn: Draft manuscript preparation, Calculations, Analysis and interpretation of results
M. M. Sharma: Draft manuscript preparation, Measurement, Analysis and interpretation of results
Dr. V. P. S. Awana: Study conception and design, Supervision, Manuscript Review
All authors reviewed the results and approved the final version of the manuscript.

**Table 1.**

| Cell Parameters |
| --- |
| Cell type: Tetragonal |
| Space Group: P4/nmm (129) |
| Refined Lattice parameters: |
| a=b=3.789(7)Å |
| & c=5.930(1) Å |
| α=β=γ=90° |
| Cell volume: 85.134 Å$^3$ |
| Density: 3.103 g/cm$^3$ |
| $\chi^2$=1.56 |
| Atomic co-ordinates: |
| Fe1 (0.25 0.75 0.00) |
| Fe2 (0.75 0.25 0.00) |
| Te (0.75 0.75 0.72) |
| Se (0.25 0.25 0.28) |

**Figure Captions:**

**Fig. 1:** (a) Shows the first Brillouin zone and with green arrows k-path for bulk electronic band structure calculation (b) The calculated Fermi surface is shown; five surfaces correspond to five bands crossing fermi level (c) The band structure of FeSe$_{0.5}$Te$_{0.5}$ with SOC parameters and without SOC parameters. The inset shows enlarged view of shaded region. (d) Shows the orbital projected density of states indicating the major orbitals contributing for valence and conduction bands.

**Fig. 2:** Surface state spectrum of FeSe$_{0.5}$Te$_{0.5}$ showing the presence of TDS cone above the Fermi level.

**Fig. 3:** (a) Rietveld refined PXRD pattern of FeSe$_{0.5}$Te$_{0.5}$ in which the red symbols show the raw data, and the black curve shows the fitting results, the left-hand inset shows the image of grown single crystals, while the unit cell of FeSe$_{0.5}$Te$_{0.5}$ is shown in middle inset, and the right-hand inset is showing the XRD pattern taken on the mechanically crystal flake. (b) Left image is the SAED pattern of FeSe$_{0.5}$Te$_{0.5}$ and right image shows the stacking of (0 0 2) planes. (c) SEM image and (d) EDAX spectra of synthesized FeSe$_{0.5}$Te$_{0.5}$ single crystal. The inset table of (d) shows the elemental composition of constituent elements of FeSe$_{0.5}$Te$_{0.5}$. (e) The recorded Raman spectroscopy of FeSe$_{0.5}$Te$_{0.5}$ at room temperature.

**Fig. 4:** (a) MR% vs. H plot of FeSe$_{0.5}$Te$_{0.5}$ for different temperatures viz. 15, 20,30, 40 K in a magnetic field range of ±14 T in which inset is showing the $\rho$-$T$ plot confirming the presence of superconductivity in FeSe$_{0.5}$Te$_{0.5}$. (b) $\rho$ vs. $\theta$ curve in a tilt angle range from 0º to 360º at 15 K under the magnetic field of 14 T, the plot is fitted by equation $|A + B\cos\theta|$ as shown by solid black curve, the inset shows the geometrical representation of the arrangement used to perform angle dependent magneto transport measurements. (c) MR% vs. H plots for different tilt angles at 15 K (d) $\Delta\sigma_{\theta,B}$ vs. $H\cos\theta$ plot in low magnetic field range, which is fitted with HLN equation shown by the solid black curve.

**References:**

1. Sato, M. and Ando, Y., Rep. Prog. Phys. **80**, 076501 (2017).
2. Leijnse, M. and Flensberg, K., Semicond. Sci. Technol. **27**, 124003 (2012).
3. Sharma, M.M., Sharma, P., Karn, N.K. and Awana, V.P.S., Supercond. Sci. Technol. **35**, 083003(2022).
4. Sato, M. and Fujimoto, S., J. Phys. Soc. Jpn. **85**, 072001 (2016).
5. Ge, J.F., Liu, Z.L., Liu, C., Gao, C.L., Qian, D., Xue, Q.K., Liu, Y. and Jia, J.F., Nat. Mater. **14**, 285 (2015).
6. Hao, N. and Hu, J., Phys. Rev. X **4**, 031053 (2014).
7. Wang, Z.F., Zhang, H., Liu, D., Liu, C., Tang, C., Song, C., Zhong, Y., Peng, J., Li, F., Nie, C. and Wang, L., Nat. Mater. **15**, 968 (2016).
8. Wang, Z., Zhang, P., Xu, G., Zeng, L.K., Miao, H., Xu, X., Qian, T., Weng, H., Richard, P., Fedorov, A.V. and Ding, H., Phys. Rev. B **92**, 115119 (2015).
9. Johnson, P.D., Yang, H.B., Rameau, J.D., Gu, G.D., Pan, Z.H., Valla, T., Weinert, M. and Fedorov, A.V., Phys. Rev. Lett. **114**, 167001 (2015).
10. Xu, G., Lian, B., Tang, P., Qi, X.L. and Zhang, S.C., Phys. Rev. Lett. **117**, 047001 (2016).
11. Mascot, E., Cocklin, S., Graham, M., Mashkoori, M., Rachel, S. and Morr, D.K., Commun. Phys. **5**, 188 (2022).
12. Wu, X., Chung, S.B., Liu, C. and Kim, E.A., Phys. Rev. Research **3**, 013066(2021).
13. Wang, D., Kong, L., Fan, P., Chen, H., Zhu, S., Liu, W., Cao, L., Sun, Y., Du, S., Schneeloch, J. and Zhong, R., Science **362**, 333 (2018).
14. Machida, T., Sun, Y., Pyon, S., Takeda, S., Kohsaka, Y., Hanaguri, T., Sasagawa, T. and Tamegai, T., Nat. Mater. **18**, 811 (2019).
15. Zhang, P., Yaji, K., Hashimoto, T., Ota, Y., Kondo, T., Okazaki, K., Wang, Z., Wen, J., Gu, G.D., Ding, H. and Shin, S., Science **360**, 182 (2018).
16. Zaki, N., Gu, G., Tsvelik, A., Wu, C. and Johnson, P.D., Proc. Natl. Acad. Sci. USA **118**, e2007241118 (2021).
17. Chauhan, H., Kumar, R. and Varma, G.D., Supercond. Sci. Technol. **35**, 045003 (2022).
18. Maheshwari, P.K., Jha, R., Gahtori, B. and Awana, V.P.S., AIP Advances **5**, 097112 (2015).
19. Cieplak, M.Z., Zajcewa, I., Lynnyk, A., Kosyl, K.M. and Gawryluk, D.J., arXiv:2211.15189 (2022).
20. Bezusyy, V.L., Gawryluk, D.J., Malinowski, A. and Cieplak, M.Z., Transition-metal substitutions in iron chalcogenides. Phys. Rev. B, **91**, 100502(2015).
21. Bezusyy, V.L., Gawryluk, D.J., Berkowski, M. and Cieplak, M.Z., arXiv:1205.6295 (2012).
22. Takagaki, Y., Giussani, A., Perumal, K., Calarco, R. and Friedland, K.J., Phys. Rev. B **86**, 125137 (2012).
23. Li, H., Wang, H.W., Li, Y., Zhang, H., Zhang, S., Pan, X.C., Jia, B., Song, F. and Wang, J., Nano Lett. **19, 4**, 2450 (2019).
24. Shiomi, Y. and Saitoh, E., AIP Advances **7**, 035011 (2017).
25. Wang, C.M. and Lei, X.L., Phys. Rev. B **86**, 035442 (2012).
26. Bao, L., He, L., Meyer, N., Kou, X., Zhang, P., Chen, Z.G., Fedorov, A.V., Zou, J., Riedemann, T.M., Lograsso, T.A. and Wang, K.L., Sci Rep **2**, 726 (2012).
27. Bhardwaj, V., Pal, S.P., Varga, L.K., Tomar, M., Gupta, V. and Chatterjee, R., Sci Rep **8**,

28. Ren, Z., Taskin, A.A., Sasaki, S., Segawa, K. and Ando, Y., Phys. Rev. B **82**, 241306(R) (2010).
29. Wang, Z., Fu, Z.G., Wang, S.X. and Zhang, P., Phys. Rev. B **82**, 085429 (2010).
30. Licciardello, S., Maksimovic, N., Ayres, J., Buhot, J., Čulo, M., Bryant, B., Kasahara, S., Matsuda, Y., Shibauchi, T., Nagarajan, V. and Analytis, J.G., Phys. Rev. Res., **1**, 023011 (2019).
31. Wang, Q., Zhang, W., Chen, W., Xing, Y., Sun, Y., Wang, Z., Mei, J.W., Wang, Z., Wang, L., Ma, X.C. and Liu, F., 2D Materials, **4**, 034004 (2017).
32. Amigó, M.L., Crivillero, V.A., Franco, D.G. and Nieva, G., J. of Phys.: Conf. Series, 568, 022005 (2014).
33. P. Giannozzi, S. Baroni, N. Bonini, M. Calandra, R. Car, C. Cavazzoni, D. Ceresoli, G. L. Chiarotti, M. Cococcioni, I. Dabo, A. Dal Corso, S. De Gironcoli, S. Fabris, G. Fratesi, R. Gebauer, U. Gerstmann, C. Gougoussis, A. Kokalj, M. Lazzeri, L. Martin-Samos, N. Marzari, F. Mauri, R. Mazzarello, S. Paolini, A. Pasquarello, L. Paulatto, C. Sbraccia, S. Scandolo, G. Sclauzero, A. P. Seitsonen, A. Smogunov, P. Umari, and R. M. Wentzcovitch, J. Phys. Condens. Matter **21**, 395502 (2009).
34. P. Giannozzi, O. Andreussi, T. Brumme, O. Bunau, M. Buongiorno Nardelli, M. Calandra, R. Car, C. Cavazzoni, D. Ceresoli, M. Cococcioni, N. Colonna, I. Carnimeo, A. Dal Corso, S. De Gironcoli, P. Delugas, R. A. Distasio, A. Ferretti, A. Floris, G. Fratesi, G. Fugallo, R. Gebauer, U. Gerstmann, F. Giustino, T. Gorni, J. Jia, M. Kawamura, H. Y. Ko, A. Kokalj, E. Kücükbenli, M. Lazzeri, M. Marsili, N. Marzari, F. Mauri, N. L. Nguyen, H. V. Nguyen, A. Otero-D Roza, L. Paulatto, S. Poncé, D. Rocca, R. Sabatini, B. Santra, M. Schlipf, A. P. Seitsonen, A. Smogunov, I. Timrov, T. Thonhauser, P. Umari, N. Vast, X. Wu, and S. Baroni, J. Phys. Condens. Matter **29**, 465901 (2017).
35. Perdew, J. P., Burke, K., & Ernzerhof, M. Generalized gradient approximation made simple. Phys. Rev. Lett. **77**, 3865 (1996).
36. Monkhorst, H. J., and Pack, J. D., Phys. Rev. B **13**, 5188 (1976).
37. Mostofi, A.A., J. R. Yates J. R., Pizzi, G., Lee, Y. S., Souza, I., Vanderbilt, D., Marzari, N., Comput. Phys. Commun. **185**, 2309 (2014).
38. Wu, Q., Zhang, S., Song, H.F., Troyer, M. and Soluyanov, A.A., Comp. Phys. Commun. **224**, 405 (2018).
39. Sancho, M.L., Sancho, J.L. and Rubio, J., Journal of Physics F: Metal Physics **14**, 1205 (1984).
40. Sancho, M.L., Sancho, J.L. and Rubio, J., Journal of Physics F: Metal Physics **15**, 851 (1985).
41. Grimme, S., Antony, J., Ehrlich, S. and Krieg, H., J. of Chem. Phys., **132** (2010).
42. Karn, N.K., Sharma, M.M. and Awana, V.P.S., Supercond. Sci. Technol., **35**, 114002 (2022).
43. Malick, S., Singh, J., Laha, A., Kanchana, V., Hossain, Z. and Kaczorowski, D., Phys. Rev. B **105**, 045103 (2022).
44. Hsu, F.C., Luo, J.Y., Yeh, K.W., Chen, T.K., Huang, T.W., Wu, P.M., Lee, Y.C., Huang, Y.L., Chu, Y.Y., Yan, D.C. and Wu, M.K., Proc. Natl. Acad. Sci. USA **105 (38)**, 14262 (2008).
45. Maheshwari, P.K., Joshi, L.M., Gahtori, B., Srivastava, A.K., Gupta, A., Patnaik, S.P. and Awana, V.P.S., Mater. Res. Express **3**, 076002 (2016).
46. Lopes, C.S., Foerster, C.E., Serbena, F.C., Júnior, P.R., Jurelo, A.R., Júnior, J.L.P., Pureur, P. and Chinelatto, A.L., Supercond. Sci. Technol., **25**, 025014 (2012).
47. Maheshwari, P.K., Reddy, V.R., Gahtori, B. and Awana, V.P.S., Mater. Res. Express, **6**, 046003 (2019).

48. Xia, T.L., Hou, D., Zhao, S.C., Zhang, A.M., Chen, G.F., Luo, J.L., Wang, N.L., Wei, J.H., Lu, Z.Y. and Zhang, Q.M., Phys. Rev. B, **79**, 140510 (2009).
49. Rayees A. Zargar, Anand Pal, A. K. Hafiz & V. P. S. Awana, J. Supercond. Nov. Magn. **27**, 897 (2014).
50. Gopal, R.K., Sheet, G. and Singh, Y., Sci. Rep. **11**, 12618 (2021).
51. Maurya, V.K., Patidar, M.M., Dhaka, A., Rawat, R., Ganesan, V. and Dhaka, R.S., Phys. Rev. B **102**, 144412 (2020).
52. Sun, Y., Taen, T., Yamada, T., Pyon, S., Nishizaki, T., Shi, Z. and Tamegai, T., Phys. Rev. B, **89**, 144512 (2014).
53. Chauhan, H., Miglani, S.K., Mitra, A. and Varma, G.D., Phys. Scr., **98**, 065934 (2023).
54. Chen, J., Qin, H.J., Yang, F., Liu, J., Guan, T., Qu, F.M., Zhang, G.H., Shi, J.R., Xie, X.C., Yang, C.L. and Wu, K.H., Phys. Rev. Lett. **105**, 176602 (2010).
55. Tang, H., Liang, D., Qiu, R.L. and Gao, X.P., ACS Nano **5**, 7510 (2011).
56. Wang, Y.Y., Yu, Q.H. and Xia, T.L., Chinese Phys. B **25**, 107503 (2016).
57. Li, H., Wang, H.W., Li, Y., Zhang, H., Zhang, S., Pan, X.C., Jia, B., Song, F. and Wang, J., Nano Lett. **19**, 2450 (2019).
58. Shrestha, K., Chou, M., Graf, D., Yang, H.D., Lorenz, B. and Chu, C.W., Phys. Rev. B **95**, 195113(2017).
59. Hikami, S., Larkin, A.I. and Nagaoka, Y., Prog. of Theo. Phys. **63**, 707 (1980).
60. Wang, J., Powers, W., Zhang, Z., Smith, M., McIntosh, B.J., Bac, S.K., Riney, L., Zhukovskyi, M., Orlova, T., Rokhinson, L.P. and Hsu, Y.T., Nano letters, **22**, 792 (2022).

**Fig. 1(a)**     **Fig. 1(b)**

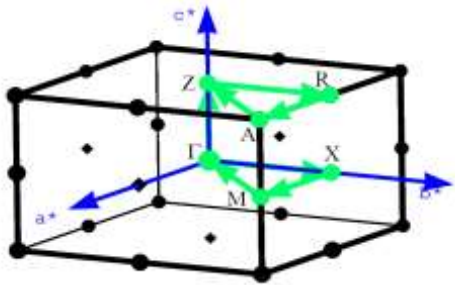   **Fig.**  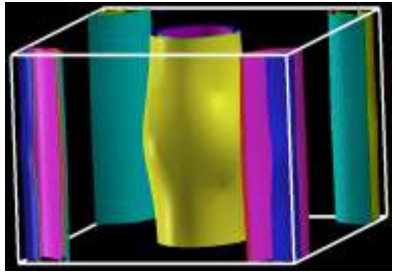

**1(c)**

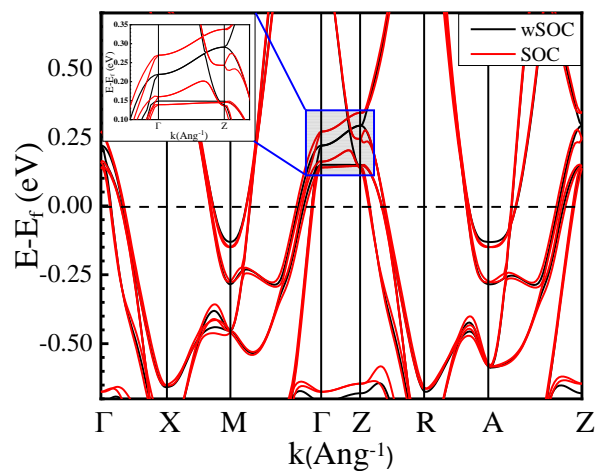

**Fig. 1(d)**

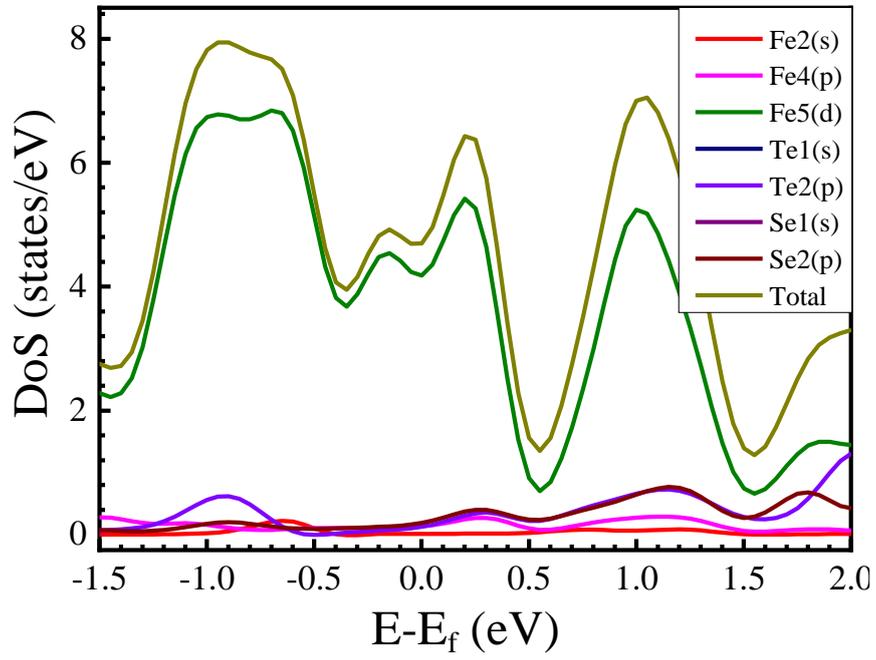

**Fig. 2**

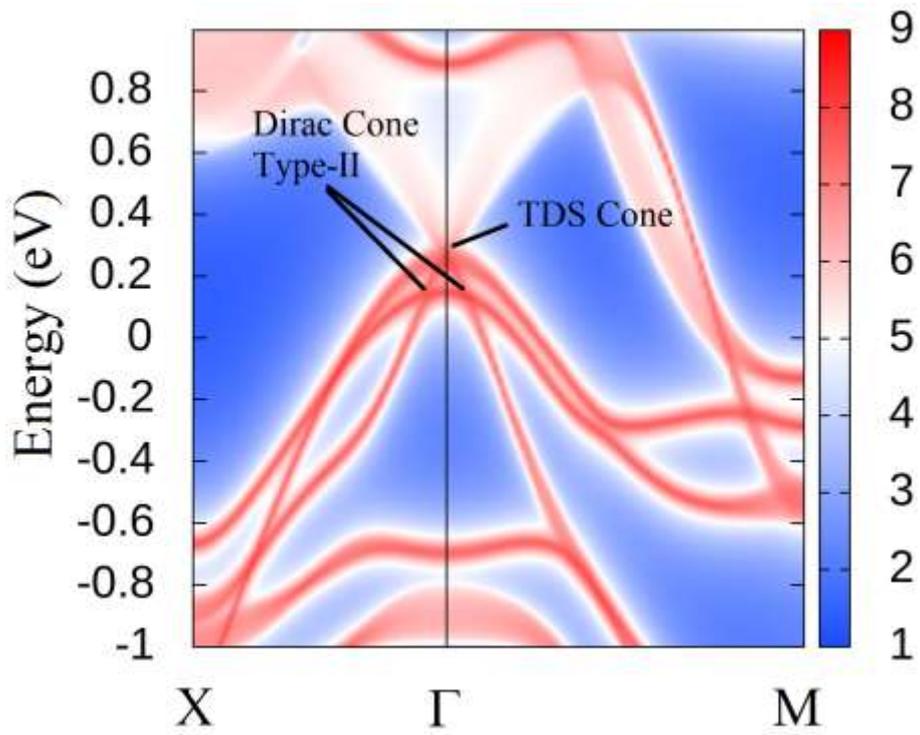

**Fig. 3(a)**

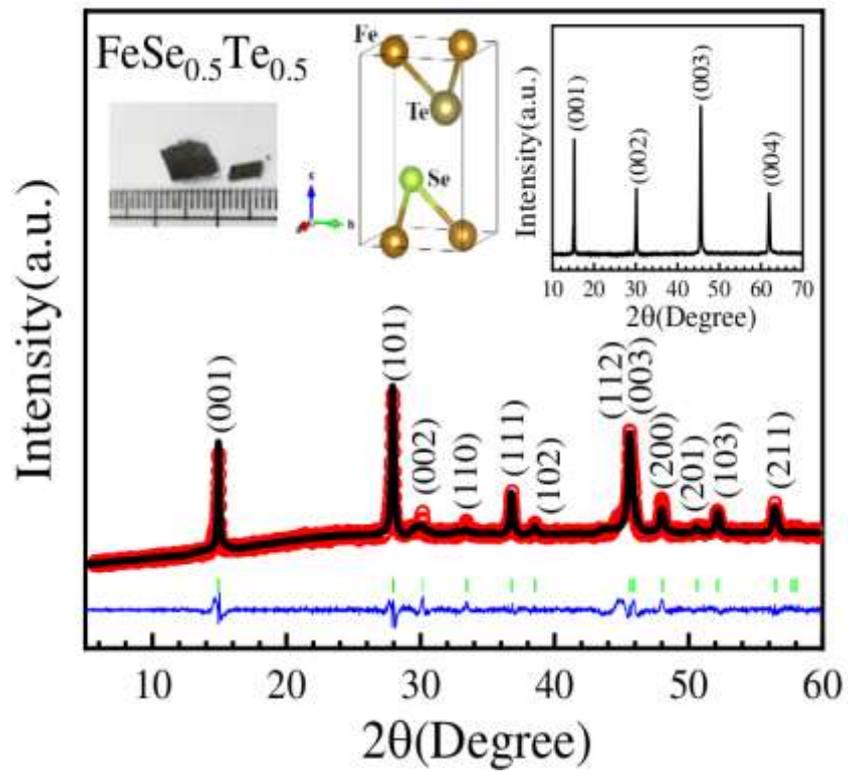

**Fig. 3(b)**

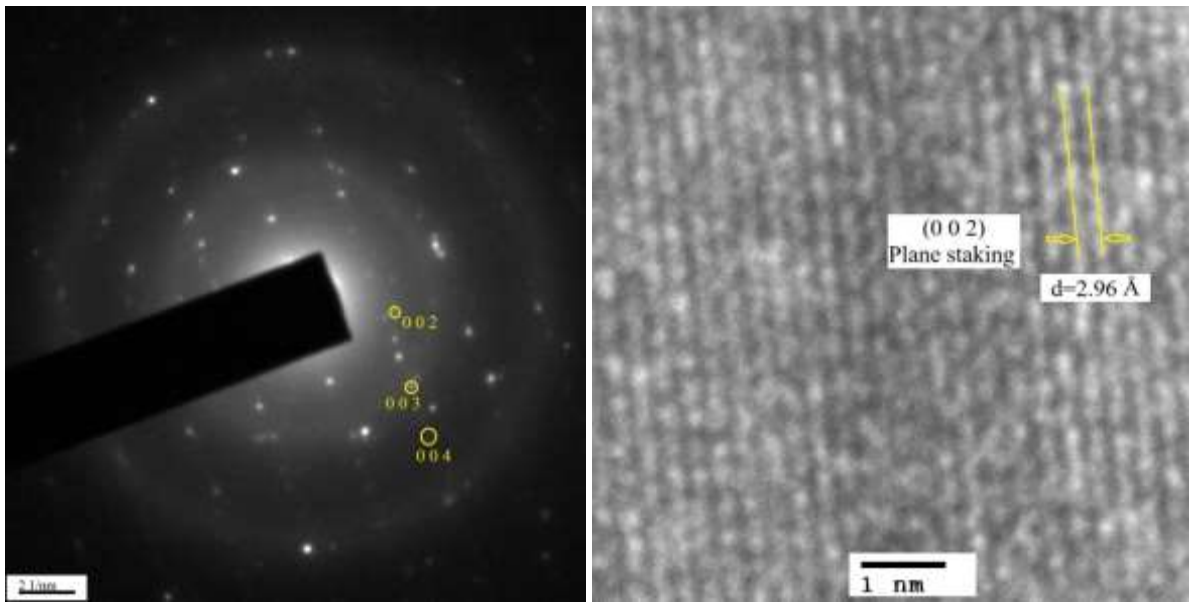

**Fig. 3(c)**

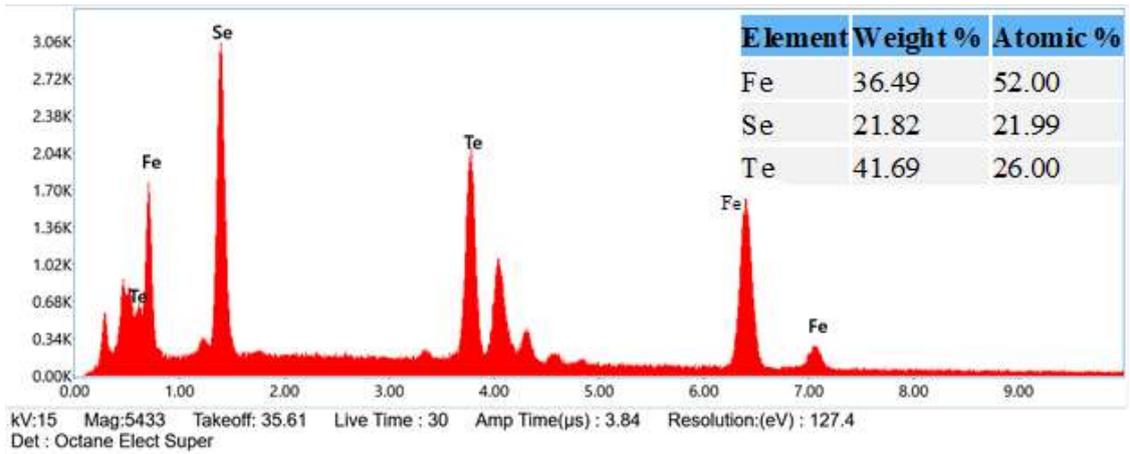

**Fig. 3(d)**

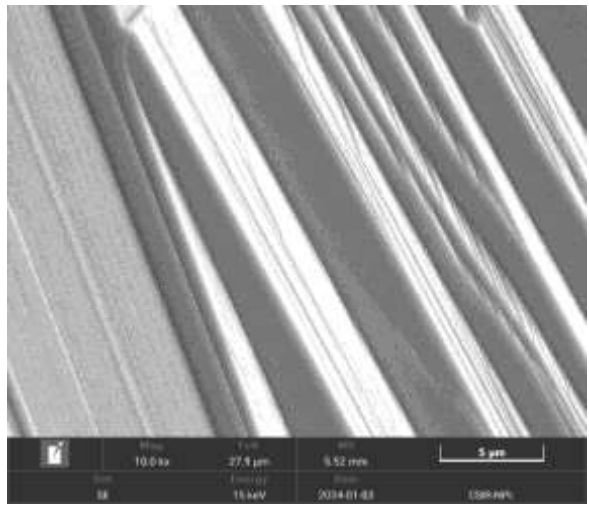

**Fig. 3(e)**

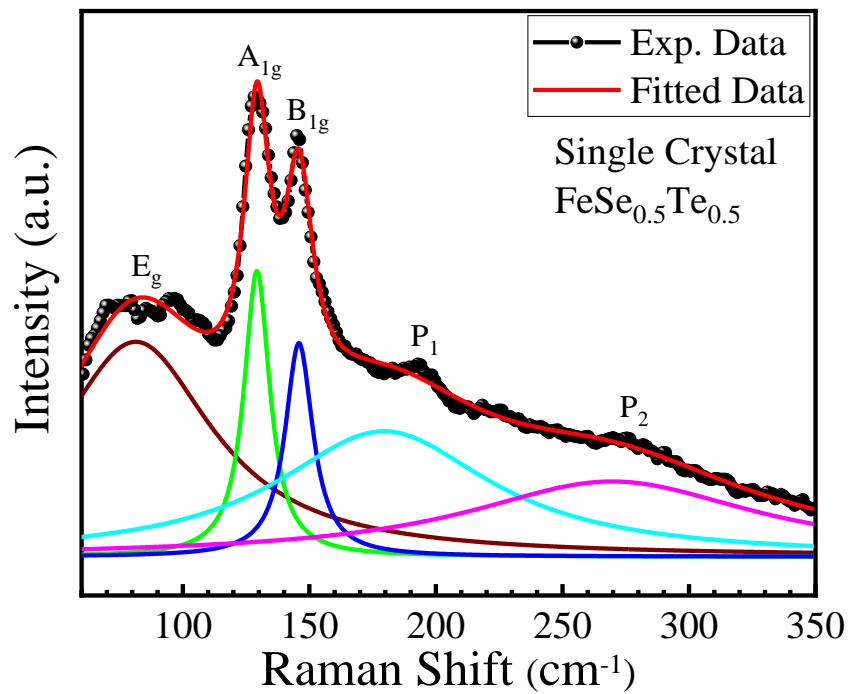

**Fig. 4(a)** 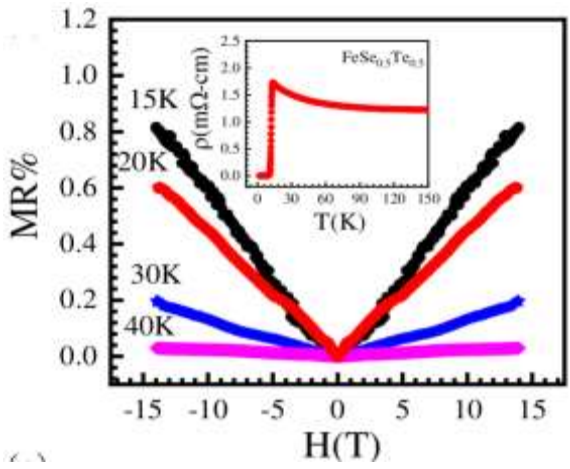

**Fig. 4(b)** 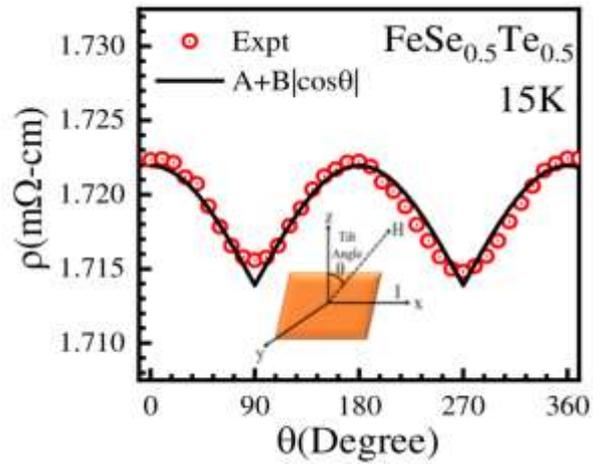

**Fig. 4(c)** 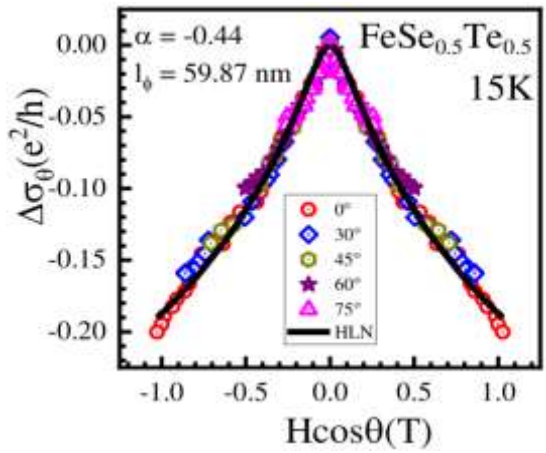

**Fig. 4(d)** 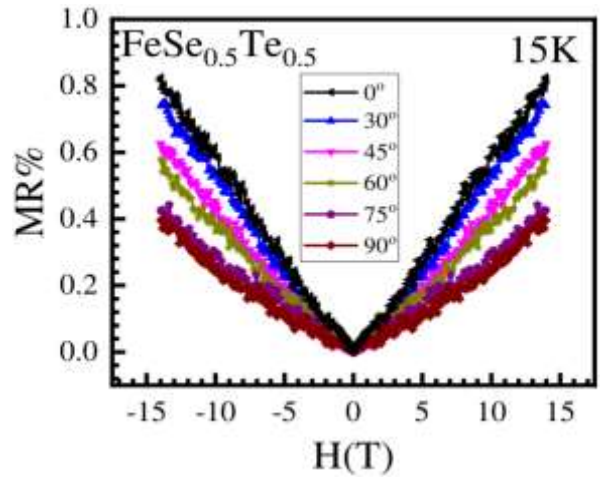